\documentclass{aipproc}

\usepackage{graphicx} 
\usepackage{amssymb}
\usepackage{amsmath}

\layoutstyle{6x9}
\usepackage{latexsym}

\begin{document}
\setlength{\parindent}{0pt}
\title{Towards a realistic axion star}
\author{J. Barranco}{
address={Instituto de F\'{\i}sica, Universidad Nacional Aut\'onoma de M\'exico, 
Apdo. Postal 20-360, 01000. M\'exico D.F. , Mexico
}
}
\author{A. Bernal}{
address={Instituto de F\'{\i}sica de la Universidad de Guanajuato, 
C.P. 37150, Le\'on, Guanajuato, Mexico
}
}

\begin{abstract}
In this work we estimate the radius and the mass of a self-gravitating system 
made of axions. The quantum axion field satisfies the Klein-Gordon equation in 
a curved space-time and the metric components of this space-time are solutions 
to the Einstein equations with a source term given by the vacuum expectation 
value of the energy-momentum operator constructed from the axion field. As a 
first step towards an axion star we consider the up 
to the $\phi^6$ term in the 
axion potential expansion. We found that axion stars would have masses of the order of 
asteroids ($\sim 10^{-10}M_\odot$) and radius of the order $\sim \mbox{few} 
~$centimeters.
\end{abstract}
\maketitle
{\bf 1. Boson stars\\}
Boson stars (BS) are gravitationally bounded systems made of scalar particles. 
They were introduced for the very first time in 1968 by Kaup and later by 
Ruffini and Bonazzola \cite{Ruffini:1969qy}. The inclusion of a 
self-interacting term was done in \cite{Colpi:1986ye} and the stability under 
general perturbations was studied in \cite{Seidel:1990jh}. BS are fully 
characterized by the scalar field properties, i.e. the mass $m$ of the 
scalar field and its potential $V(\phi)=m^2\phi^2+\lambda\phi^4/2$, where 
$\lambda$ is the self-interaction parameter. BS rise as solutions of the 
Einstein-Klein-Gordon equations
\begin{equation}\label{einstein}
G_{\mu\nu}=8 \pi G < T_{\mu\nu} >\,,
\qquad \left(\Box - \frac{dV}{d\phi^2}\right)\phi=0\,,
\end{equation}
where $\Box=(1/\sqrt{-g})\partial_\mu[\sqrt{-g}g^{\mu \nu}\partial_\nu]$ 
and the energy-momentum tensor is given by 
\begin{equation}\label{tensor}
T_{\mu\nu}=\partial_{\mu}\Phi \partial_{\nu}\Phi-\frac{1}{2}g_{\mu \nu}(
g^{\alpha \beta}\partial_{\alpha}\Phi \partial_{\beta}\Phi+m^2\Phi^2+\frac{\lambda}{2}\Phi^4)\,.
\end{equation}
Here $<...>$ denotes average over the ground state of the system of many particles. The field $\phi$ is second quantized. In this formalism, $\phi=\phi^{+}+\phi^{-}$ is an operator, where we have defined 
$\phi^{+}=\sum\mu_{lmn}^+R_{nl}(r)Y^l_m(\theta ,\psi)e^{-iE_nt}$ ,
$\phi^{-}=\sum\mu_{lmn}^-R_{nl}(r)Y^{l*}_m(\theta, \psi)e^{+iE_nt}$ and $\mu_{lmn}^\pm$ are the creation (annihilation) operators for a particle with angular momentum $l$, azimuthal momentum $m$ and energy $E_n$. From operator $\phi$ it is possible to construct the energy-momentum tensor $\langle Q|T_{\mu \nu}|Q\rangle$ by considering an state $|Q \rangle$ for which all the $N$ particles are in the ground state ($l=m=0$, $n=1$). 
It is assumed an static, spherical symmetric space-time metric
\begin{equation}\label{metric}
ds^2=-B(r)dt^2+A(r)dr^2+r^2d\Omega^2 \,.
\end{equation}
By demanding regular solution at the origin and flatness at infinity, there exists a family of radial solutions characterized by the number of nodes in the radial coordinate. Those with zero nodes are called ground states. That solutions exhibit a maximum mass for a specific initial central field value
$\phi_c(0)$. It is known that boson stars with initial central field value $\phi(0) < \phi_c(0)$ are stable and those with $\phi(0) > \phi_c(0)$ are unstable. The magnitude of the maximum mass of the boson star depends on the value of the self-interacting coupling. It was shown in  \cite{Colpi:1986ye} that 
maximum mass scales  as $M_{max}=0.22 \Lambda^{1/2}M_p^2/m$ where $M_p$ is the Planck mass, $\Lambda=\lambda M_p^2 / 4 \pi m^2$ is the adimensional self-interacting coupling and $m$ the mass particle associated with the scalar field.\\  

{\bf 2. The axion\\} 
An axion is a scalar field originated at the  Peccei-Quinn symmetry breaking  phase transition \cite{Peccei:1977hh}. It is one of the favorite candidates to account for the dark matter content of the universe. Its properties have been strongly constrained from astrophysical and cosmological considerations. The axion decay constant is allowed to be in the range $10^{10} \mbox{GeV} \le f_a \le 10^{12}\mbox{GeV}$ and its mass in the range $10^{-5} \mbox{eV} \le m \le 10^{-3}\mbox{ev}$.
At late stages of the evolution of the universe it acquires an non-vanishing potential energy density
$V(\phi)=m^2 f_a^2 \left[ 1 - \mbox{cos}\left( {\phi \over f_a} \right)
\right]\,.$\\

{\bf 3. Towards an axion star\\}
By doing and expansion of $\mbox{cos}(\phi/f_a)$ 
we obtain
\begin{equation}\label{axionp}
V(\phi) \sim {1 \over 2} m^2 \phi^2-{1\over 4!}m^2 {\phi^4\over f_a^2}+
{1\over 6!}m^2 {\phi^6 \over f_a^4}-...
\end{equation}

The addition of the potential energy density (\ref{axionp}) to the kinetic 
energy of axions give rise to a energy-momentum tensor of energy similar to eq. 
(\ref{tensor})
except for a sign with the proper identification $\lambda=m^2/6 f_a^2$.
It is natural to investigate if this energy density gives rise to standard boson
stars as those studied in \cite{Colpi:1986ye}. We proceed by following 
\cite{Ruffini:1969qy} with the potential energy eq. (\ref{axionp}). The Einstein-Klein-Gordon  equations (\ref{einstein}) obtained with the metric (\ref{metric}) and $\langle Q|T_{\mu \nu}|Q\rangle$ computed as explained above are  
\begin{eqnarray}\label{sistema}
{A'\over A^2 x}+{1 \over x^2}\left(1-{1 \over A}\right)-
\left[\left({1 \over B}+1\right)\sigma^2-{\Lambda \over 2}\sigma^4+
{\sigma'^2 \over A}+{\Lambda^2 \over 10} \sigma^6 \right]&=&0 \,, \nonumber \\
{B'\over ABx}-{1 \over x^2}\left(1-{1\over A} \right)-
\left[\left({1\over B}-1\right)\sigma^2+{\Lambda \over 2}\sigma^4+
{\sigma'^2 \over A}-{\Lambda^2 \over 10}\sigma^6 \right] &=&0 \,, \nonumber\\
\sigma''+\left({1\over x}+{B'\over 2B}-{A' \over 2A}\right)\sigma'+
A\left[ \left({1\over B}-1\right)\sigma+\Lambda\sigma^3-{3\over 10}\Lambda^2
\sigma^5\right]&=&0\,,
\end{eqnarray}
where the prime denotes derivative with respect to $x$ and we have done the definitions $x=rm$, $R_{10}={1 \over \sqrt{4 \pi G}}\sigma$, $B=E_{1}^2B$ and 
$\Lambda={1\over 24\pi}\left( {m_p \over f_a} \right)^2\,.$
\begin{center}
\begin{figure}
\includegraphics[angle=270,width=0.6\textwidth]{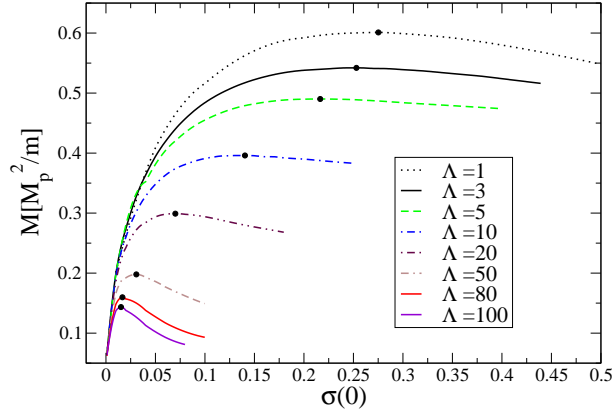}
\caption{Mass of a boson star as a function of the central value of the scalar field. Circles denote the critical point, that is, the maximum mass of a boson star. The stars with a central value to the right of this point are unstable and those to the left are stable.}\label{masses}
\end{figure}
\end{center}
{\bf 4. Results\\}

We have solved the system (\ref{sistema}) for different values of $\Lambda$. As in the $\lambda |\phi|^4$ case, we found a maximum mass for specific $\sigma(0)=\sigma_c(0)$, but, the switch in the potential sign produces a significant change in the behavior of the maximum mass. The relation $M_{max}\sim \Lambda^{1/2}$ is not satisfied anymore. Instead of increasing $M_{max}$ we found decreasing $M_{max}$ as can be easily observe in Fig. (\ref{masses}). We have extrapolated the maximum mass as a function of $\Lambda$ 
\begin{equation}\label{extra1}
M_{max}=M_{max}(\Lambda)=-{0.844\over \Lambda}+{1.396\over \Lambda^{1/2}}+{0.0450\over \Lambda^{1/4}}\,.
\end{equation}
Plugging the axion values of $f_a$ and $m$ in $\Lambda$ we found $\Lambda_{axion} \sim 10^{13}-10^{17}$ and using this value in extrapolated $M_{max}$ eq. (\ref{extra1}) we found $10^{-12} M_\odot < M_{max}(\Lambda_{axion}) < 10^{-10}M_\odot \,$. The value $\sigma_c(0)$, where the maximum mass is reached, depends strongly on $\Lambda$. Again, by doing an extrapolation, we can estimate the critical density for and axion star. Another interesting result is that the radius doesn't depend on $\Lambda$ but it depends on $\sigma(0)$. Once we have estimated $\sigma_c(0)$ for axion star we can then estimate the radius using the extrapolation. We found $\sigma_c(\Lambda_{axion})\sim 10^{-4}$ and the adimensional radius 
$R(\sigma_c(\Lambda_{axion})) \sim 10^2$, that is,  $R \sim$ few cm. for axion mass $m \sim 10^{-3}$ eV.

{\bf Acknowledgments} A.B. has been partially supported by SEP-2004-C01-47641. 
J.B. is partially supported by DGAPA-UNAM.



\end{document}